# High-sensitivity COVID-19 group testing by digital PCR

Authors: Alexandra Martin (1), Alexandre Storto (2,3), Barbara André (1), Allison Mallory (1), Rémi Dangla (1), Benoit Visseaux (2,3), Olivier Gossner (4)

((1) Stilla Technologies, Villejuif, France, (2) Université de Paris, Assistance Publique – Hôpitaux de Paris, Service de virologie, Hôpital Bichat, Paris, France (3) UMR 1137-IAME, DeSCID: Decision SCiences in Infectious Diseases control and care, INSERM, Université de Paris, Paris, France, (4) CREST, Ecole Polytechnique, Palaiseau, France)

Comments: 17 pages, 6 figures

## Abstract

Background: Worldwide demand for SARS-CoV-2 RT-PCR testing is increasing as more countries are impacted by COVID-19 and as testing remains central to contain the spread of the disease, both in countries where the disease is emerging and in countries that are past the first wave but exposed to re-emergence. Group testing has been proposed as a solution to expand testing capabilities but sensitivity concerns have limited its impact on the management of the pandemic. Digital PCR (RT-dPCR) has been shown to be more sensitive than RT-PCR and could help in this context.

Methods: We implemented RT-dPCR based COVID-19 group testing on commercially available system and assay (Naica™ System from Stilla Technologies) and investigated the sensitivity of the method in real life conditions of a university hospital in Paris, France, in May 2020. We tested the protocol in a direct comparison with reference RT-PCR testing on 448 samples split into groups of 3 sizes for RT-dPCR analysis: 56 groups of 8 samples, 28 groups of 16 samples and 14 groups of 32 samples.

Results: Individual RT-PCR testing identified 25 positive samples. Using groups of 8, testing by RT-dPCR identified 23 groups as positive, corresponding to 26 true positive samples including 2 samples not initially detected by individual RT-PCR but confirmed positive by further RT-PCR and RT-dPCR investigation. For groups of 16, 15 groups tested positive, corresponding to 25 true positive samples identified. 100% concordance is found for groups of 32 but with limited data points.

Conclusions: Our proposed approach of group testing by digital PCR is shown to have a similar to better diagnostic sensitivity compared to individual RT-PCR testing for group sizes of up to 16 samples. This approach reduces the quantity of reagent needed by up to 80% while reducing costs and increasing capabilities of testing by up to 10-fold.

# Introduction

As the first wave of the COVID-19 is fading in most countries of the Northern hemisphere, many of them have started implementing extensive monitoring policies to prevent the apparition of new clusters potentially leading to a second wave. These policies all require important testing capabilities, as was exemplified in Wuhan where all 11 Million citizens were recently tested in 10 days during May 2020. In some Southern countries, the amount of new cases is still increasing and wide population testing cannot be achieve everywhere. Thus, scaling up and maintaining large testing capacities worldwide remains a challenge, with limited reagent production and scarcity of testing equipment likely to remain limitations.

Group testing or pooling, first suggested by Dorfman in 1943, is a protocol through which individual samples are combined together before running the test(1). The advantage of the method is an overall saving in the number of tests required to screen a given population (2), and thereby an increase in testing capabilities for fixed reagent and instrumentation availability. Savings depend on key parameters such as the disease prevalence and the group size. Group testing protocols using real-time reverse-transcriptase PCR (RT-PCR) have been evaluated and implemented for Covid-19 screening around the world in several experiments using RT-PCR detection techniques, notably in Israel, Germany, California, Nebraska, NY State, and Italy (3–9).

Although these studies show that positive individuals can be detected in pooled samples, the number of amplification cycles needed to detect (Ct value) those is often increased by dilution and perhaps by inhibition effects (3,5,7,9). This can prevent weakly positive specimen from being detected in group samples (3,8). Concerns about the sensitivity of group testing have been raised by French medical authorities, leading to a negative recommendation on their use in France(10).

Digital PCR (or RT-dPCR) is a PCR technique known for its higher sensitivity and precision over classical RT-PCR (11,12). Digital PCR has also been shown to be more resistant to PCR inhibitors (13). Recent studies have confirmed high sensitivity of RT-dPCR for the detection of SARS-CoV-2 (14–16).

In this study we propose a novel group testing protocol using a commercially available RT-dPCR assay and compare empirically the sensitivity of individual identification through RT-PCR with group testing by RT-dPCR for three groups sizes of 8, 16 and 32 samples. We find that, in our condition of evaluation, group testing by RT-dPCR has a better or similar sensitivity than the reference individual RT-PCR testing for groups of 8 and 16.

# Material and Methods

## Summary of the method of the comparative study

Overall, 448 patient samples are tested for SARS-CoV-2 by i) individual RT-PCR (local gold standard method), ii) RT-dPCR in 56 groups of 8 samples, iii) RT-dPCR in 28 groups of 16 samples and iv) RT-dPCR in 14 groups of 32 samples and results are compared between all four test protocols. In case of discordance between the results of individual RT-PCR testing and group testing in RT-dPCR, samples were re-analyzed individually by RT-dPCR, by the gold-standard RT-PCR and a confirmatory RT-PCR assay. The whole protocol is illustrated in Figure 1.

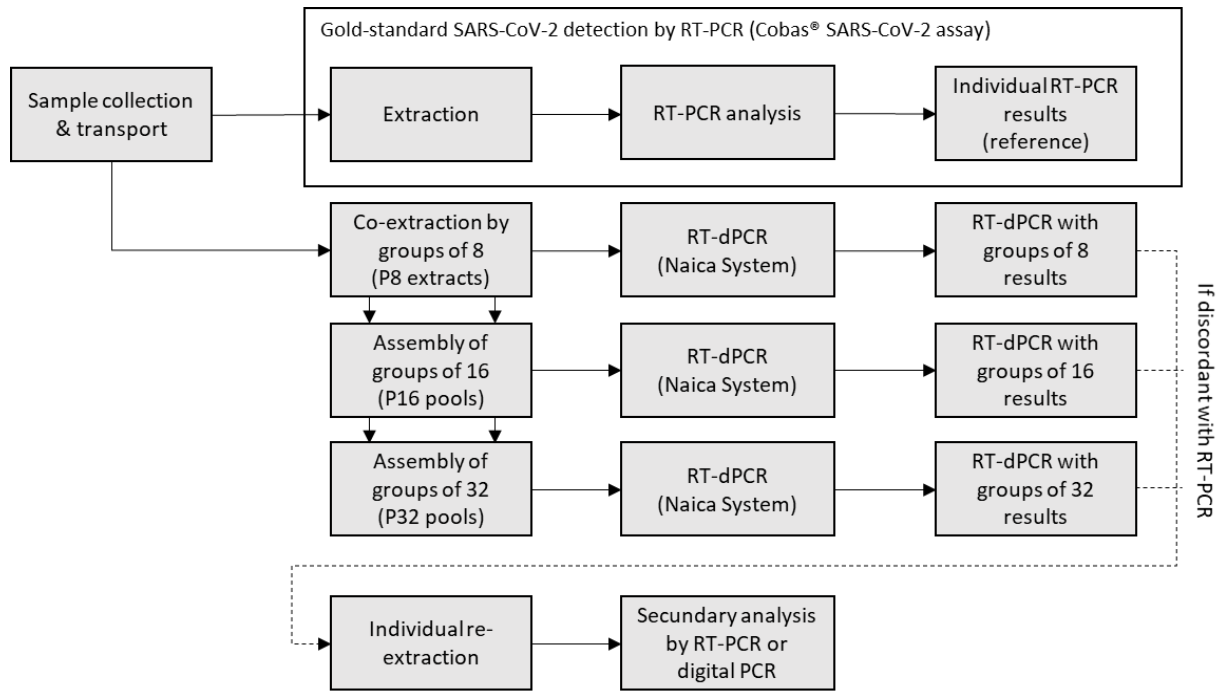

*Figure 1 : Schematic of the structure of the comparative study.*

### Specimens collection, storage and pooling

Nasopharyngeal swabs of 448 patients screened for Covid-19 at the Bichat university hospital (Paris, France), one of the two Paris reference centers for emerging diseases, between May 6[th] and May 26[th], 2020 were included. All samples were collected in universal transport medium (UTM) (Virocult® or eSwab[TM]) and tested, within 15 hours maximum upon collection, for SARS-CoV-2 detection according to the local standards. Briefly, 400 µL of transport medium were tested by RT-PCR (Cobas SARS-CoV-2 test, Roche). All the remaining volume of transport medium of all specimens were kept at +5°C.

No later than 24 hours after routine screening, all samples with a leftover UTM volume above 600 µL were systematically included in the group testing analysis. Thus, 125 µL of each included specimen were sampled and randomly mixed with seven others to generate a total of 56 groups of 8 specimens with a final volume of 1 mL per group. Nucleic acids were extracted from each group prior to viral titration by RT-dPCR. The remaining volume of transport medium was stored at +5°C for further investigations if required.

According to the current French ethical laws, samples used in the current study were only included after the completion of all analysis required for the patient's care.

### Detection of SARS-CoV-2 by routine individual RT-PCR testing

All 448 specimens were analyzed individually on a Cobas® 6800 system (Roche, Switzerland) for Covid-19 screening using the Cobas® SARS-CoV-2 Test kit targeting conserved regions for *ORF-1a/b* and *E-gene*. For each specimen, 400 µL of transport medium were mixed with 400 µL of Cobas® lysis buffer and loaded on the robot. During the run, extracts were eluted in 50 µL of which 27 µL were used in the RT-PCR amplification of *E* and *ORF-1a/b*. A sample was considered positive for routine screening of COVID-19 ("RT-PCR+") if either target had a Ct value below 40 PCR cycles.

Within 11 days maximum (and 20 days for "Sample_25659") upon storage at +5°C, some samples which had different results for RT-PCR and RT-dPCR were reassessed on the Cobas® 6800 system. To compensate for the low remaining amounts of transport medium, the nasal swabs were vortexed once more into the remaining transport medium diluted 1 to 10 with new transport medium.

### Extraction of total NAs on grouped samples

All nucleic acids extractions for RT-dPCR assays were performed on a MagNA Pure LC 2.0 (Roche, Switzerland) using the MagNA Pure LC Total Nucleic Acid Isolation Kit (Roche, Switzerland) following manufacturer's instructions. For all sample groups, the total volume of 1 mL was used for the extraction. For individual samples, 200 µL was diluted with 800 µL of buffer before extraction. Nucleic acids were eluted from 1mL to 50 µL of the elution buffer provided with the kit and stored at +5°C for a maximum of 12 hours before analysis.

### Preparation of groups of 16 and 32 individuals

After extraction of the 56 groups of 8 specimen (P8 extracts) and prior to viral titration by RT-dPCR, 28 groups of 16 individual samples (P16 groups) were obtained by mixing 15 µL of 2 P8 extracts and 14 groups of 32 (P32 groups) were obtained by mixing 10 µL of 2 P16 groups.

### Detection of SARS-CoV-2 by grouped RT-dPCR testing

SARS-CoV-2 titration of the grouped samples by RT-dPCR was performed on the Naica system (Stilla Technologies, France) within the next three hours after extraction, using the COVID-19 Multiplex Digital PCR Detection Kit (Stilla Technologies, France/Apexbio, China). This one-step reverse-transcription PCR kit is a triplex PCR allowing amplification, detection and quantification of one sequence in the *N* gene, one sequence in the *ORF1ab* region of SARS-CoV2 and an endogenous internal control (IC) to assess the quality of the sample and extraction. These sequences are targeted by three TaqMan probes respectively labelled with a FAM, HEX and Cy®5 fluorophore.

As recommended by the kit manufacturer, the PCR mix for a single reaction contained 12.5 µL of dPCR MasterMix 1, 1 µL of dPCR Mix 2, 1 µL of COVID-19 Assay and 10.5 µL of either, P8, P16, P32, positive control, negative control or individual extract. The 25 µL of this PCR mix were loaded in the inlet ports of the Sapphire chips (Stilla Technologies, France). The chips were placed in the Naica Geode (Stilla Technologies, France) for droplets generation, reverse transcription and PCR amplification following the kit manufacturer's instructions.

After amplification, the chips were transferred to the Naica Prism3 (Stilla technologies, France) for fluorescence reading in the three detection channels and data were analyzed with Crystal Miner Software (Stilla Technologies, France) following the kit manufacturer's instructions.

An illustration of the resulting fluorescence dot-plots used to quantify the SARS-CoV-2 virus by RT-dPCR is shown in Supplementary Materials.

### Individual confirmatory testing for SARS-CoV-2 by RT-PCR and RT-dPCR

In case of discrepancies between individual RT-qPCR and grouped RT-dPCR, RT-dPCR results were confirmed by extracting individually each sample of the group and retesting them individually according to the previously described RT-dPCR protocol. Confirmatory RT-qPCR test were also performed on individual samples as previously described with the Cobas SARS-CoV-2 assay and a third method, RealStar® SARS CoV-2 RT-PCR Kit (Altona Diagnostics, Germany), was also performed on an ABI 7500 thermocycler (ThermoFisher Scientific, United-States) from the same individual extraction as for the confirmatory RT-dPCR assay. Briefly, this latter assay targets all lineage B-betacoronaviruses by amplifying a sequence of the *E* gene and a sequence of the *S* gene specific to SARS-CoV-2 as well as a heterologous internal control. A sample was considered positive to SARS-CoV-2 if the Ct value of either target is below 40 PCR cycles.

## LoB/LoD evaluation for SARS-CoV-2 detection using RT-dPCR

The Limit of Blank (LoB) and Limit of Detection (LoD) were evaluated for SARS-CoV-2 detection using the group testing approach used in the study on a cohort of 256 pre-epidemic nasal swab samples (negative control samples) that were collected between December 1st 2019 and January 31st 2020 and for which transport medium was stored at -20 °C within 48h after sampling.

Specimens were randomly grouped into 32 groups of 8 negative controls which were co-extracted and analyzed by RT-dPCR using the same protocol described above. The results for all 32 groups are given in Supplementary Materials. The LoB at 95% confidence level for the *N* target and the *ORF1ab* target is determined to be of 2 and 0 positive droplets respectively. The LoD at 95% confidence level for each target is of 0.49 copies / µL (7 copies/PCR) and 0.24 copies/µL (3 copies/PCR), respectively.

Consequently, a threshold of a least 3 positive droplets in aggregate between both the *N* target and *ORF1ab* target is used to classify a sample as positive to SARS-CoV-2 by RT-dPCR in this study.

# Results

## Cohort description from routine RT-PCR testing

Using routine RT-PCR testing, 25 samples were identified as positive out of the 448 samples tested, corresponding to an average test positivity rate of 5.5%. The positivity rate decreased during the study period from 8.5% for the first 224 samples to 2.5% for the last 224 samples, in correlation with the decreasing disease prevalence observed during the month of May in France.

The average Ct value was of 30.0 and 27.3 for the *E* gene and *ORF* gene respectively, with minimum values of 16.5 and 16.3 and maximum values of 38.7 and >40 (not detected) (Figure 2).

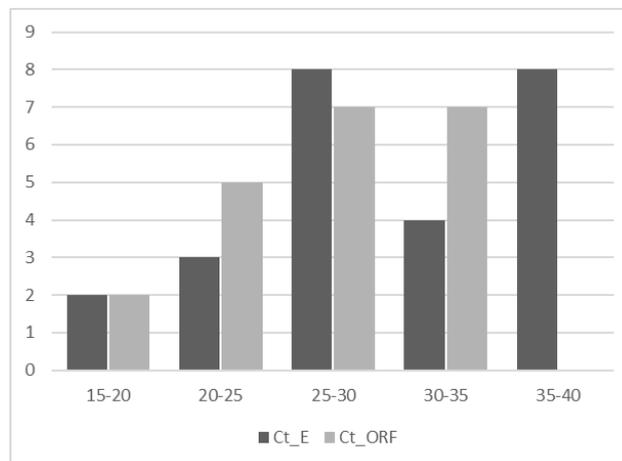

*Figure 2 : Distribution of Ct values for the E gene and ORF gene, as measured using individual reference RT-PCR with Cobas® 6800 SARS-CoV-2 assay, for the 25 positive samples.*

## Results from grouped RT-dPCR testing

All results for the detection of SARS-CoV-2 by RT-dPCR for the grouped extracts (P8) and the subsequent groups of 16 (P16) and 32 (P32) are presented in Table 1. Detailed results are listed in Supplementary Materials.

| Number of RT-PCR positive sample(s) in the group | Results for P8 extracts | | | Results for P16 groups | | | Results for P32 groups | | |
|---|---|---|---|---|---|---|---|---|---|
| | Total | dPCR - | dPCR + | Total | dPCR - | dPCR + | Total | dPCR - | dPCR + |
| 0 | 35 | 32 | 3 | 12 | 11 | 1 | 2 | 2 | 0 |
| 1 | 18 | 1 | 17 | 10 | 2 | 8 | 6 | 0 | 6 |
| 2 | 2 | 0 | 2 | 4 | 0 | 4 | 3 | 0 | 3 |
| 3 | 1 | 0 | 1 | 1 | 0 | 1 | 1 | 0 | 1 |
| 4 | 0 | 0 | 0 | 1 | 0 | 1 | 1 | 0 | 1 |
| 5 | 0 | 0 | 0 | 0 | 0 | 0 | 0 | 0 | 0 |
| 6 | 0 | 0 | 0 | 0 | 0 | 0 | 1 | 0 | 1 |
| Total | 56 | 33 | 23 | 28 | 12 | 16 | 14 | 2 | 12 |

*Table 1: Distribution of the samples identified as positive by the routine RT-PCR method (Cobas® SARS-CoV-2 assay) in the groups of 8, 16 and 32 and corresponding RT-dPCR detection of SARS-CoV-2 in the groups.*

Because sample pooling was performed systematically as samples came in the laboratory for routing RT-PCR testing, the groups contain variable numbers of RT-PCR positive samples ("RT-PCR+" samples).

Given the positivity rate of RT-PCR at the time of the study, the majority of P8 extracts had no RT-PCR+ samples (35 out of 56) and 21 groups contained at least one RT-PCR+ sample, including 18 that had one single RT-PCR+ sample.

For the largest group size of 32 samples, only 2 P32 groups had no RT-PCR+ samples. 12 P32 groups had at least one RT-PCR+ sample, including 6 with only one single RT-PCR+ sample.

### Detailed results for RT-dPCR in groups of 8

The results for SARS-CoV-2 detection by RT-dPCR in groups of 8 samples, detailed in Tables 1 and 2, are in concordance with the reference individual RT-PCR testing for 52 groups (corresponding for 416 samples), out of which 32 were RT-PCR negative groups and 20 groups contained at least one RT-PCR+ sample.

For the remaining 4 groups with discording results, three RT-PCR negative groups tested positive by RT-dPCR ("RT-PCR-/dPCR+" discordances – group IDs: P8_20, P8_28 and P8_39) and one RT-PCR+ group was found negative by RT-dPCR (RT-PCR+/dPCR- discordance – group ID: P8_02). The Ct values for the sample associated with the RT-PCR+/dPCR- discordance was of 34 and 32.3 for the *E* gene and *ORF1ab* with the Cobas® SARS-CoV-2 assay, respectively. This sample is referred to as *Sample 25659* for later discussion.

Complementary analysis of the discordances are depicted below.

Of note, out of the 8 individual samples with Ct > 35 for the *E* gene, 6 ended up to be the only positive sample in a P8 extract and all the corresponding groups were detected positive by RT-dPCR. The highest detected Ct values for the *E* gene and *ORF1ab* were 38.7 and >40 (not detected), respectively.

| Confusion matrix for P8 extracts | Expected negatives (RT-PCR) | Expected positives (RT-PCR) | Total |
|---|---|---|---|
| Negatives in RT-dPCR | 32 | 1 | 33 |
| Positives in RT-dPCR | 3 | 20 | 23 |
| Total | 35 | 21 | 56 |

*Table 2: Confusion matrix for P8 extracts*

### Detailed results for RT-dPCR in groups of 16

The results for SARS-CoV-2 detection by RT-dPCR in groups of 16 samples, detailed in Tables 1 and 3, are in concordance with individual RT-PCR testing for 25 groups (corresponding for 400 samples), out of which 11 are RT-PCR- groups and 14 are RT-PCR+ groups. Among the three groups with discording results, one presented a RT-PCR-/dPCR+ discordance (Group ID: P16_14) and two RT-PCR+/dPCR- discordances (Group IDs: P16_13 and P16_28).

Of note, out of the 8 individual samples with Ct > 35 for the *E* gene, 5 ended up to be the only positive sample in a P16 group. Two of these groups are responsible for the 2 RT-PCR+/dPCR- discordances. The *E* gene and *OFR1ab* Ct values for these 2 samples were of [36.7; >40 (not detected)] and [36.3; 34.2], while the highest Ct values for a detected single positive sample were [38.3; >40 (not detected)].

| Confusion matrix for P16 groups | Expected negatives (RT-PCR) | Expected positives (RT-PCR) | Total |
|---|---|---|---|
| Negatives in RT-dPCR | 11 | 2 | 13 |
| Positives in RT-dPCR | 1 | 14 | 15 |
| Total | 12 | 16 | 28 |

*Table 3: Confusion matrix for P16 groups*

Detailed results for RT-dPCR in groups of 32

The results for SARS-CoV-2 detection by RT-dPCR in groups of 32 are in concordance with individual RT-PCR testing for all 14 groups (corresponding for 448 samples) and are depicted in Tables 1 and 4.

Out of the 8 individual samples with Ct >35 for the *E* gene, 3 ended up to be the only positive sample in a P32 group. All such 3 P32 groups tested positive by RT-dPCR. The highest detected Ct values for the *E* gene and *ORF1ab* is of [36.7; >40 (not detected)].

| Confusion matrix for P32 groups | Expected negatives (RT-PCR) | Expected positives (RT-PCR) | Total |
|---|---|---|---|
| Negatives in RT-dPCR | 2 | 0 | 2 |
| Positives in RT-dPCR | 0 | 12 | 12 |
| Total | 2 | 12 | 14 |

*Table 4: Confusion matrix for P32 groups*

### Investigation of RT-PCR-/dPCR+ discordances

As detailed above, three P8 extracts (P8_20, P8_28 & P8_39) and one P16 group (P16_14) tested positive by RT-dPCR while containing only RT-PCR negative samples. P16_14 originates from a combination of groups with opposite discordances: P8_27 (RT-PCR+/dPCR-) and P8_28 (RT-PCR-/dPCR+) groups.

To further investigate these RT-PCR-/dPCR+ discordances, confirmatory testing RT-dPCR was performed on all individual samples from corresponding groups. For each group, one sample tested positive by individual RT-dPCR, with measured concentrations of viral RNA ranging from 128 copies per reaction to 2 copies per reaction for the *N* gene, and from 106 to 1 copies for the *ORF1ab* gene. The three corresponding dPCR+ samples (sample IDs: 52042, 56075 and 60401) were retested on the Cobas® 6800 system and by confirmatory individual RT-qPCR using the Altona assay. All results are given in the Supplementary Materials.

Two samples were found positive using the Altona assay with Ct values ranging between 28.4 and 33 for the *E* and *S* genes. Among them, the sample presenting the highest viral load by RT-dPCR (Sample 56075) was also found positive by the Cobas® confirmatory assay with a high Ct value of 36.7 for the *E* gene while the *ORF* gene was not detected. The remaining sample tested negative with both the confirmatory Cobas assay and the Altona assay. It had borderline levels of positive droplets in RT-dPCR (N=2; ORF1ab=1).

Based on these results and for further sensitivity discussions, samples *52042* and *56075* that tested positive by both RT-dPCR and Altona RT-PCR are considered as true positive samples. Sample 60401 is considered an RT-dPCR false positive pending further analysis.

### Investigation of the sample RT-PCR+/dPCR-

One group of 8 (P8_02) was tested negative by RT-dPCR and contained one RT-PCR+ sample (*Sample 25659*) with Ct values of 34 and 32.3 for the *E* gene and *ORF1ab* respectively. *Sample 25659* was subsequently retested by RT-PCR on the Cobas protocol and was also re-extracted individually and retested by individual RT-dPCR.

*Sample_25659* was found to be borderline negative by RT-dPCR (N=2; ORF1ab=0) but Ct values of 37.3 and 34.9 were found for *E* and *ORF* respectively in the second Cobas® assessment.

Based on these results and for further sensitivity discussions, sample 25659 is considered as a true positive samples.

### Correlation between RT-dPCR measurements and Ct values

A good correlation was observed between the number of positive droplets observed in RT-dPCR and the Ct values of the positive samples contained in the groups. See Supplementary Materials for more details.

## Discussion

In this work, we assessed the sensitivity and specificity of group testing combined with digital PCR for SARS-CoV-2 detection. Three different group sizes were investigated using a commercially available digital PCR assay, the COVID-19 Multiplex Digital PCR Detection Kit (Stilla Technologies, France/Apexbio, China). This assay demonstrated a low LoB (at 2 and 0 positive droplets per PCR for *N* and *ORF1ab* genes, respectively) and low LoD (at 5 and 3 copies/PCR for *N* and *ORF1ab* genes, respectively). This LoD is lower than most estimation for WHO and other reference RT-PCR assays typically ranging between 5 to 500 copies/PCR (17,18).

For our analysis, we proposed a protocol of group screening performed by RT-dPCR with secondary individual re-testing of positive groups as illustrated in Figure 3.

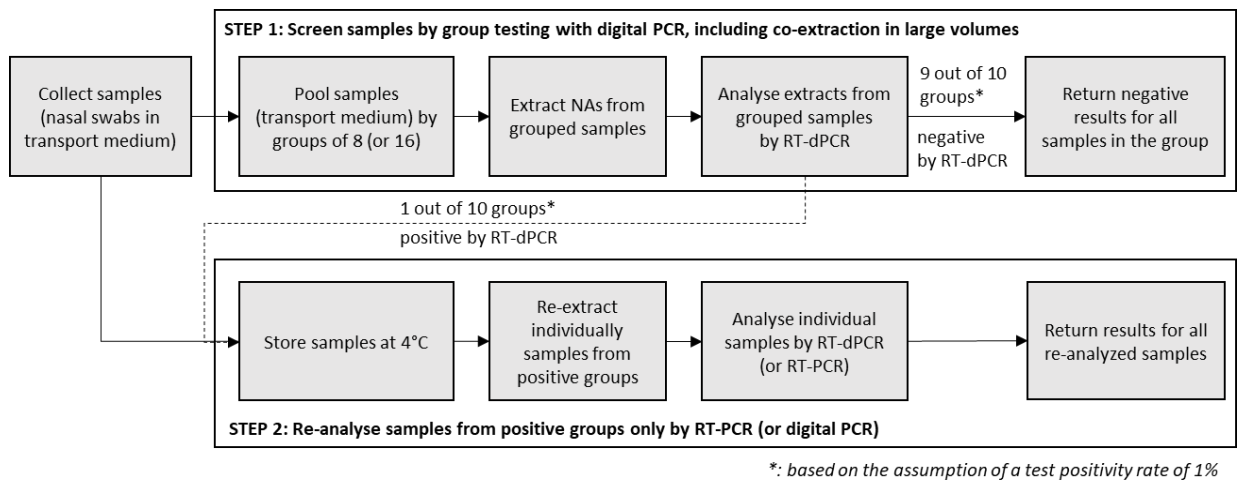

*: based on the assumption of a test positivity rate of 1%*

*Figure 3 : Graph of the suggested practical protocol for implementation of group testing by RT-dPCR.*

We assessed this protocol by assessing in real-life condition 448 consecutive samples grouped by 8, 16 and 32 samples.

We observed a better sensitivity than individual RT-PCR testing for groups of 8 samples. Indeed, a total of 23 groups of 8 samples tested positive and included 26 true positive samples. Only 25 samples were identified as positive using reference individual RT-PCR testing, corresponding to an +4% improvement in sensitivity. Two among the three samples associated with RT-PCR-/dPCR+ discordances were confirmed as true positive by RT-PCR using a SARS-CoV-2 specific assay from Altona. The last one was not confirmed as positive by RT-PCR and is undergoing further investigations. One sample was associated to a RT-PCR+/dPCR- discrepant results and is also considered a true positive sample as confirmed by RT-PCR retesting. Further investigation by sequencing is underway on these 4 samples in an effort to rigorously assess the existence and nature of nucleic acids variations responsible for the discording results.

Grouped testing by RT-dPCR has similar sensitivity to individual RT-PCR testing for a group size of 16 samples. 15 groups of 16 samples tested by RT-dPCR positive and included a total of 25 true positive samples, of which 23 are RT-PCR positive samples and 2 are RT-PCR-/dPCR+ samples. However, 2 RT-PCR+ groups tested negative with grouped RT-dPCR. This is likely explained by high Ct values of the single positive sample included in each of these 2 groups.

Testing in the 14 groups of 32 samples by RT-dPCR has 100% concordance with the reference RT-PCR testing. Re-testing positive groups by RT-dPCR would have likely led to better sensitivity than RT-PCR. However, we are careful in drawing conclusions for groups of 32 given the limited data points in this case as only 14 groups, including only 2 RT-PCR negative groups.

An alternative and even more cost-effective group testing protocol could be to perform the re-testing steps using RT-PCR with Cobas or Altona assays. In these protocols, the sensitivity becomes dependent on the RT-PCR kit used, leading to potential discrepancies with RT-dPCR as observed in our results for groups of 8 samples. However, our data suggest that performing the individual tests for groups of 8 and 16 with a RT-PCR assay would still have a similar sensitivity.

Overall, our data indicates that COVID-19 group testing combined with digital PCR for large group sizes of 8 and 16 samples, followed by individual re-testing of positive groups, has better or similar sensitivity than individual RT-PCR testing. Groups of 32 samples could also be considered, but our analysis needs to be confirmed in the future as, due to the epidemiological situation at the time, a low number of negative groups and groups containing a single positive sample were assessed.

The gain in sensitivity of the proposed method is likely due to a combination of *i)* a concentration effect due to performing the pooling prior extraction and performing the extraction step from a large volume of 1 mL of pooled transport medium and *ii)* the intrinsic superior sensitivity of digital PCR compared to RT-PCR, as demonstrated previously for SARS-CoV-2 (14–16) and other viruses (12,19) detection.

Below standard sensitivity is one of the main reasons why group testing has not been widely adopted for COVID-19 testing, whilst research groups have advocated for its implementation as a solution to the world-wide un-met demand for tests and reagent shortage(2-9).

The current study suggests that the high sensitivity of group testing could be achieved using by digital PCR instead of RT-PCR in the first group screening step. It makes the approach viable for large-scale, low cost patient screening with minimum reagent consumption. For a test positivity rate of 1%, our group testing protocol would save 79%, 80% and 69% of reagents compared to individual testing, for group sizes of 8, 16 and 32 samples. Group size does not significantly impact cost or reagent consumption for group sizes above 8 samples and test positivity rates below 1%. Consequently, the choice of group size (8, 16 or 32) is mostly a balance of sensitivity and test capacity. Our results indicate that group testing by digital PCR and a group size of 16 would increase testing capabilities by more than 10-fold while having a sensitivity comparable to the current standard of individual testing by RT-PCR.

Group testing can be used in various context where testing is not currently put in place due to testing capacity or with strong economics constraints and where SARS-CoV-2 prevalence is low. In countries where the pandemic is not yet under control or could re-emerge, enhancing testing capacity is essential to winning the battle against COVID-19. In countries where the first wave is fading, the fight against COVID-19 goes through a combination between a test / trace / isolate strategy and social distancing. Increasing the range of people tested amongst contacts with positive cases, but also periodic testing of population in frequent contact with others (e.g. nurses, transportation workers, clerks, etc…) as well as in fragile populations such as nursing homes can be part of a strategy against COVID-19 while allowing a relaxation of social distancing measures at the same time. Group testing can help in all of these situations.

# Supplementary Materials

## S1 – Data format for RT-dPCR as obtained with the Naica System

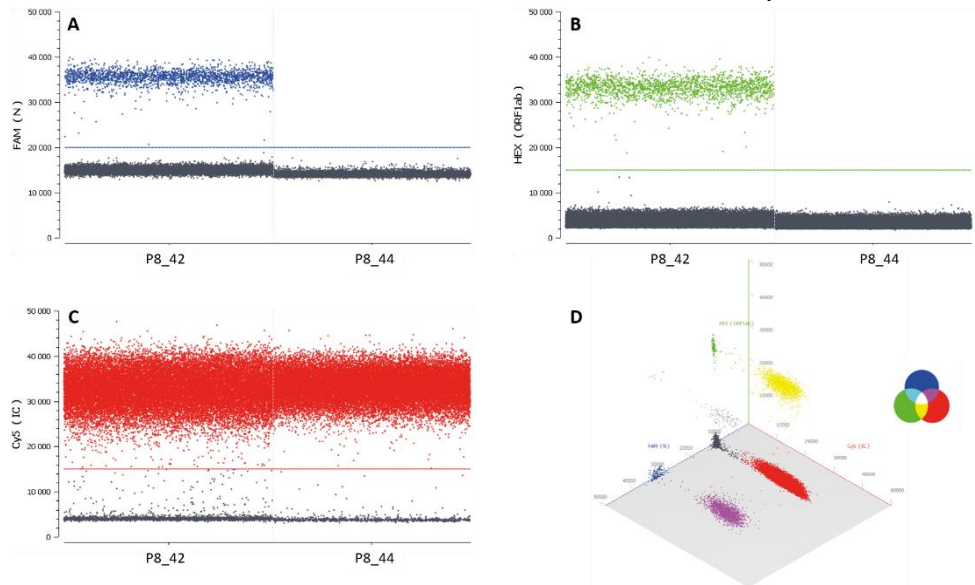

*Figure S1.1: Fluorescence results as displayed in the Crystal Miner software showing 1D dot-plots of a positive (P8_42) and a negative (P8_44) pool of 8 samples in the FAM (A), HEX (B) and Cy5 (C) channels. The horizontal line marks the threshold above which droplets (represented as dots) are considered positive for the amplification of N, ORF1ab and for the endogenous internal control respectively. The thresholds are set by manufacturer at 20 000 RFUs, 15 000 RFUs and 15 000 RFUs for the FAM, HEX and Cy5 channels. (D) 3D dot-plots of P8_42: if the concentration of either target is high enough, co-encapsulation of several targets can occur in a droplet leading to the apparition of clusters for double positive droplets (cyan, yellow and purple) and triple positive droplets (light grey). Triple negative droplets, containing no target, are shown in the dark grey cluster.*

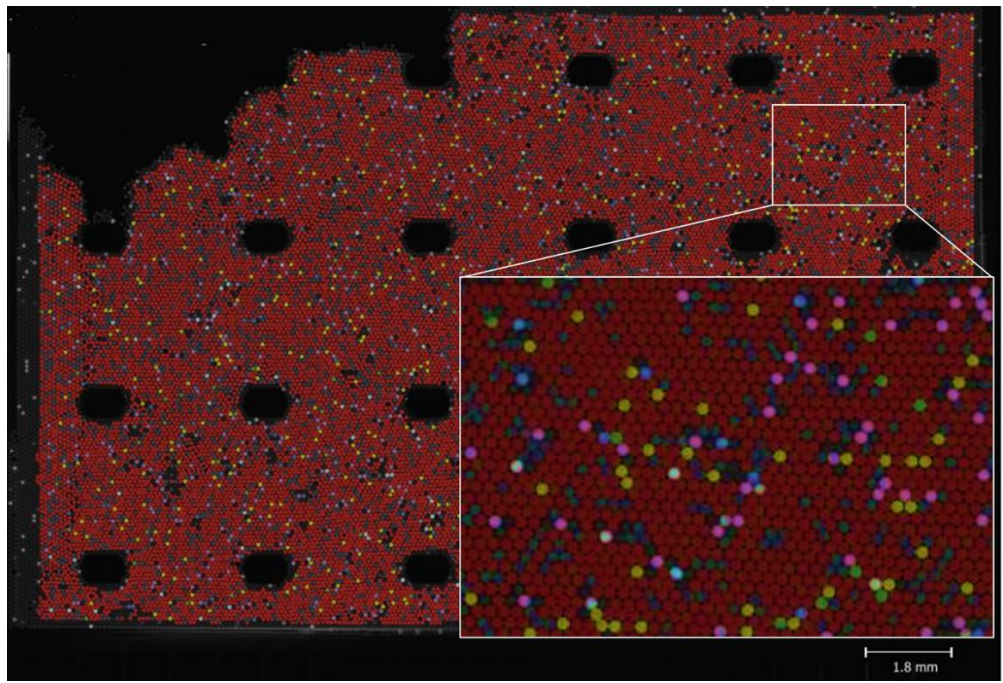

*Figure S1.2: Image of the droplet crystal obtained using the Naica System on grouped extracts P8_22, including a zoom on a sub-region of the crystal. Droplet color code: dark grey = negative for all targets ; bleu : positive for N gene only ; green : positive for ORF1ab only; red : positive for IC only; cyan : positive for N and ORF1ab genes; magenta: positive for N gene and IC; yellow: positive for ORF1ab gene and IC; white / mixed: positive for all. The droplet crystal contains 25 820 analyzable droplets, out of which 1 057, 883 and 21 121 were positive for the N, ORF1ab and IC respectively.*

## S2 – Result table for LoB & LoD calculations

| Sample ID | Number of droplets | Pos. droplets (N+ORF1ab) | Pos. droplets (N) | Pos. droplets (ORF1ab) | IC (cp/µL) |
|---|---|---|---|---|---|
| Group 1 | 19460 | 0 | 0 | 0 | 31430 |
| Group 2 | 20641 | 0 | 0 | 0 | 10985 |
| Group 3 | 21449 | 0 | 0 | 0 | 4054 |
| Group 4 | 20983 | 0 | 0 | 0 | 15143 |
| Group 5 | 19881 | 1 | 1 | 0 | 7401 |
| Group 6 | 20609 | 0 | 0 | 0 | 8783 |
| Group 7 | 19044 | 0 | 0 | 0 | 6249 |
| Group 8 | 20787 | 0 | 0 | 0 | 8759 |
| Group 9 | 21470 | 0 | 0 | 0 | 10271 |
| Group 10 | 18952 | 0 | 0 | 0 | 17578 |
| Group 11 | 23424 | 0 | 0 | 0 | 8222 |
| Group 12 | 22748 | 0 | 0 | 0 | 9875 |
| Group 13 | 23747 | 0 | 0 | 0 | 10275 |
| Group 14 | 24886 | 0 | 0 | 0 | 4266 |
| Group 15 | 24079 | 0 | 0 | 0 | 10264 |
| Group 16 | 24718 | 0 | 0 | 0 | 13219 |
| Group 17 | 24228 | 0 | 0 | 0 | 10186 |
| Group 18 | 23742 | 0 | 0 | 0 | 8431 |
| Group 19 | 23880 | 0 | 0 | 0 | 7628 |
| Group 20 | 23613 | 0 | 0 | 0 | 3111 |
| Group 21 | 25723 | 0 | 0 | 0 | 8987 |
| Group 22 | 25522 | 0 | 0 | 0 | 7411 |
| Group 23 | 24781 | 0 | 0 | 0 | 6894 |
| Group 24 | 7074 | 0 | 0 | 0 | 5884 |
| Group 25 | 27003 | 0 | 0 | 0 | 8911 |
| Group 26 | 27138 | 0 | 0 | 0 | 5552 |
| Group 27 | 24425 | 0 | 0 | 0 | 7886 |
| Group 28 | 27017 | 0 | 0 | 0 | 7484 |
| Group 29 | 26722 | 0 | 0 | 0 | 9963 |
| Group 30 | 26088 | 0 | 0 | 0 | 2492 |
| Group 31 | 26747 | 0 | 0 | 0 | 5703 |
| Group 32 | 26728 | 0 | 0 | 0 | 5825 |

*Table S2: Detection results in number of positive droplets for N, ORF1ab, N+ORF1ab and IC by RT-dPCR in 32 groups of 8 pre-epidemic samples.*

The pre-epidemic groups were used as negative controls for SARS-CoV-2 detection. Only one out of 32 (group 5) had one positive droplet for the *N* target out of 19881 droplets analyzed. The LOB at 95% confidence level for this RT-dPCR assay was estimated automatically by the Gene-Pi online statistical tool (https://www.gene-pi.com/statistical-tools/loblod/), considering a droplet volume of 0.548 nL as specified by the manufacturer. The LOD at 95% confidence level is computed from the experimental LOB. The LOB for the *N* and *ORF1ab* targets are found to be of 2 and 0 droplets respectively. The LODs are found to be of 0.46 copies/µL and 0.22 copies/µL, or equivalently of 7 and 3 copies per PCR reaction.

Using a joint analysis, a LOB analysis can be performed for the combination of the *N* and *ORF1ab* targets. In the case, the sum of the positives for the *N* and *OFR1ab* target is used as the measure and the LOB for the sum is 2 positive droplets. Consequently, in this study, a RT-dPCR result is considered as positive when the sum of positive droplets in the *N* and *ORF1ab* targets is of 3 droplets or more (strictly greater than 2).

For details on calculation models, see:
- https://www.gene-pi.com/wp-content/uploads/2018/03/Memo_LOB_calculation_method.pdf
- https://www.gene-pi.com/wp-content/uploads/2019/11/GenePi_Memo_LOD_calculation_method-1.pdf

# S3 - Full results of pooled testing by RT-dPCR

| Sample_ID | Ct_E | Ct_ORF | P8_## | # of droplets | Pos droplets N | Pos droplets ORF | IC [C] (cp/µL) | COBAS/dPCR status | P16_## | # of droplets | Pos droplets N | Pos droplets ORF | IC [C] (cp/µL) | COBAS/dPCR status | P16_## | # of droplets | Pos droplets N | Pos droplets ORF | IC [C] (cp/µL) | COBAS/dPCR status |
|---|---|---|---|---|---|---|---|---|---|---|---|---|---|---|---|---|---|---|---|---|
| 25652 | 21.2 | 20.9 | P8_01 | 25340 | 25340 | 25128 | 15556 | +/+ | P16_01 | 25262 | 25196 | 22657 | 9080 | +/+ | P32_01 | 17348 | 16047 | 11332 | 9237 | +/+ |
| 25653 | 38.01 | 0 | | | | | | | | | | | | | | | | | | |
| + 6 COBAS negative samples | | | | | | | | | | | | | | | | | | | | |
| 25659 | 34.02 | 32.26 | P8_02 | 25685 | 0 | 0 | 3642 | +/- | | | | | | | | | | | | |
| + 7 COBAS negative samples | | | | | | | | | | | | | | | | | | | | |
| 8 COBAS negative samples | | | P8_03 | 22995 | 1 | 0 | 5530 | -/- | P16_02 | 24508 | 0 | 0 | 10026 | -/- | | | | | | |
| 8 COBAS negative samples | | | P8_04 | 26245 | 0 | 0 | 14896 | -/- | | | | | | | | | | | | |
| 27241 | 34 | 32.1 | P8_05 | 26634 | 63 | 42 | 24859 | +/+ | P16_03 | 26718 | 980 | 350 | 15988 | +/+ | P32_02 | 4298 | 3890 | 3693 | 17412 | +/+ |
| + 7 COBAS negative samples | | | | | | | | | | | | | | | | | | | | |
| 27304 | 28.8 | 28.12 | P8_06 | 27184 | 1767 | 650 | 8160 | +/+ | | | | | | | | | | | | |
| + 7 COBAS negative samples | | | | | | | | | | | | | | | | | | | | |
| 27316 | 27.71 | 27.1 | P8_07 | 25752 | 3597 | 2667 | 13527 | +/+ | P16_04 | 26735 | 26497 | 26315 | 19941 | +/+ | | | | | | |
| + 7 COBAS negative samples | | | | | | | | | | | | | | | | | | | | |
| 27342 | 26.27 | 25.24 | P8_08 | 26060 | 26059 | 26056 | 28478 | +/+ | | | | | | | | | | | | |
| + 7 COBAS negative samples | | | | | | | | | | | | | | | | | | | | |
| 30923 | 38.27 | ND | P8_09 | 25655 | 4 | 1 | 16784 | +/+ | P16_05 | 26446 | 4 | 1 | 18236 | +/+ | P32_03 | 21285 | 83 | 54 | 19589 | +/+ |
| + 7 COBAS negative samples | | | | | | | | | | | | | | | | | | | | |
| 8 COBAS negative samples | | | P8_10 | 25511 | 0 | 0 | 19934 | -/- | | | | | | | | | | | | |
| 8 COBAS negative samples | | | P8_11 | 22083 | 0 | 0 | 17689 | -/- | P16_06 | 27274 | 224 | 125 | 21805 | +/+ | | | | | | |
| 25736 | 34.5 | 32 | P8_12 | 22503 | 418 | 256 | 29721 | +/+ | | | | | | | | | | | | |
| + 7 COBAS negative samples | | | | | | | | | | | | | | | | | | | | |
| 31173 | 38.73 | ND | P8_13 | 22408 | 3 | 2 | 13871 | +/+ | P16_07 | 23420 | 23420 | 23418 | 13459 | +/+ | P32_04 | 24222 | 24222 | 24221 | 14252 | +/+ |
| + 7 COBAS negative samples | | | | | | | | | | | | | | | | | | | | |
| 31271 | 18.71 | 17.81 | P8_14 | 22982 | 22982 | 22974 | 14204 | +/+ | | | | | | | | | | | | |
| 31278 | 36.52 | 34.83 | | | | | | | | | | | | | | | | | | |
| 31417 | 25.04 | 24.61 | | | | | | | | | | | | | | | | | | |
| + 5 COBAS negative samples | | | | | | | | | | | | | | | | | | | | |
| 8 COBAS negative samples | | | P8_15 | 22883 | 0 | 0 | 17210 | -/- | P16_08 | 21923 | 17166 | 14218 | 14676 | +/+ | | | | | | |
| 31397 | 21.6 | 21.16 | P8_16 | 24406 | 23431 | 21878 | 12568 | +/+ | | | | | | | | | | | | |
| 31415 | 25.04 | 24.61 | | | | | | | | | | | | | | | | | | |
| + 6 COBAS negative samples | | | | | | | | | | | | | | | | | | | | |
| 8 COBAS negative samples | | | P8_17 | 25825 | 0 | 0 | 13476 | -/- | P16_09 | 24573 | 739 | 554 | 13055 | +/+ | P32_05 | 23569 | 319 | 202 | 12648 | +/+ |
| 51318 | 28.35 | 27.89 | P8_18 | 25304 | 1543 | 1189 | 15304 | +/+ | | | | | | | | | | | | |
| + 7 COBAS negative samples | | | | | | | | | | | | | | | | | | | | |
| 51958 | 35.15 | 32.25 | P8_19 | 25088 | 40 | 39 | 16078 | +/+ | P16_10 | 26314 | 32 | 25 | 16918 | +/+ | | | | | | |
| + 7 COBAS negative samples | | | | | | | | | | | | | | | | | | | | |
| 52042 | ND | ND | P8_20 | 25083 | 5 | 2 | 14024 | -/+ | | | | | | | | | | | | |
| + 7 COBAS negative samples | | | | | | | | | | | | | | | | | | | | |
| 8 COBAS negative samples | | | P8_21 | 24808 | 0 | 2 | 18935 | -/- | P16_11 | 24624 | 468 | 360 | 12282 | +/+ | P32_06 | 26550 | 305 | 212 | 13244 | +/+ |
| 52408 | 28.81 | 28.22 | P8_22 | 25820 | 1057 | 883 | 7399 | +/+ | | | | | | | | | | | | |
| + 7 COBAS negative samples | | | | | | | | | | | | | | | | | | | | |
| 8 COBAS negative samples | | | P8_23 | 23991 | 0 | 0 | 12420 | -/- | P16_12 | 25837 | 0 | 0 | 14885 | -/- | | | | | | |
| 8 COBAS negative samples | | | P8_24 | 25510 | 0 | 0 | 16601 | -/- | | | | | | | | | | | | |

| Sample_ID | Ct_E | Ct_ORF | P8_## | # of droplets | Pos droplets N | Pos droplets ORF | IC [C] (cp/µL) | COBAS/dPCR status | P16_## | # of droplets | Pos droplets N | Pos droplets ORF | IC [C] (cp/µL) | COBAS/dPCR status | P16_## | # of droplets | Pos droplets N | Pos droplets ORF | IC [C] (cp/µL) | COBAS/dPCR status |
|---|---|---|---|---|---|---|---|---|---|---|---|---|---|---|---|---|---|---|---|---|
| 8 COBAS negative samples | | | P8_25 | 25585 | 1 | 0 | 13405 | -/- | P16_13 | 25257 | 0 | 0 | 15031 | +/- | P32_07 | 25105 | 7 | 2 | 13330 | +/+ |
| 27309 | 36.65 | ND | P8_26 | 25498 | 5 | 1 | 18210 | +/+ | | | | | | | | | | | | |
| + 7 COBAS negative samples | | | | | | | | | | | | | | | | | | | | |
| 8 COBAS negative samples | | | P8_27 | 25770 | 0 | 0 | 14721 | -/- | P16_14 | 25364 | 13 | 5 | 13146 | -/+ | | | | | | |
| 56075 | ND | ND | P8_28 | 25667 | 32 | 19 | 16170 | -/+ | | | | | | | | | | | | |
| + 7 COBAS negative samples | | | | | | | | | | | | | | | | | | | | |
| 8 COBAS negative samples | | | P8_29 | 25568 | 0 | 0 | 11813 | -/- | P16_15 | 27217 | 0 | 0 | 12192 | -/- | P32_08 | 26284 | 0 | 0 | 11980 | -/- |
| 8 COBAS negative samples | | | P8_30 | 25757 | 0 | 0 | 12565 | -/- | | | | | | | | | | | | |
| 8 COBAS negative samples | | | P8_31 | 25476 | 0 | 0 | 8230 | -/- | P16_16 | 27198 | 0 | 0 | 12741 | -/- | | | | | | |
| 8 COBAS negative samples | | | P8_32 | 25324 | 0 | 0 | 17214 | -/- | | | | | | | | | | | | |
| 8 COBAS negative samples | | | P8_33 | 25765 | 1 | 0 | 17060 | -/- | P16_17 | 26670 | 0 | 0 | 17539 | -/- | P32_09 | 26986 | 4628 | 3336 | 13335 | +/+ |
| 8 COBAS negative samples | | | P8_34 | 25803 | 0 | 0 | 17915 | -/- | | | | | | | | | | | | |
| 59120 | 24.1 | 23.28 | P8_35 | 26095 | 14584 | 11406 | 9821 | +/+ | P16_18 | 25884 | 8381 | 6405 | 11333 | +/+ | | | | | | |
| + 7 COBAS negative samples | | | | | | | | | | | | | | | | | | | | |
| 8 COBAS negative samples | | | P8_36 | 25774 | 0 | 0 | 13463 | -/- | | | | | | | | | | | | |
| 8 COBAS negative samples | | | P8_37 | 24635 | 1 | 0 | 10338 | -/- | P16_19 | 26555 | 0 | 0 | 14937 | -/- | P32_10 | 25842 | 230 | 184 | 12290 | +/+ |
| 8 COBAS negative samples | | | P8_38 | 24082 | 1 | 0 | 17996 | -/- | | | | | | | | | | | | |
| 60401 | ND | ND | P8_39 | 26374 | 1 | 3 | 15263 | -/+ | P16_20 | 27345 | 506 | 415 | 11245 | +/+ | | | | | | |
| + 7 COBAS negative samples | | | | | | | | | | | | | | | | | | | | |
| 60611 | 28.14 | 27.86 | P8_40 | 26259 | 979 | 757 | 7136 | +/+ | | | | | | | | | | | | |
| + 7 COBAS negative samples | | | | | | | | | | | | | | | | | | | | |
| 74465 | 16.5 | 16.3 | P8_41 | 26883 | 26883 | 26883 | 8581 | +/+ | P16_21 | 24825 | 24825 | 24825 | 9690 | +/+ | P32_11 | 26391 | 26391 | 26391 | 9388 | +/+ |
| + 7 COBAS negative samples | | | | | | | | | | | | | | | | | | | | |
| 75547 | 31.3 | 29.4 | P8_42 | 27389 | 2045 | 1905 | 11346 | +/+ | | | | | | | | | | | | |
| + 7 COBAS negative samples | | | | | | | | | | | | | | | | | | | | |
| 8 COBAS negative samples | | | P8_43 | 26105 | 0 | 0 | 5791 | -/- | P16_22 | 26078 | 0 | 0 | 9354 | -/- | | | | | | |
| 8 COBAS negative samples | | | P8_44 | 25839 | 0 | 0 | 14504 | -/- | | | | | | | | | | | | |
| 8 COBAS negative samples | | | P8_45 | 27194 | 0 | 1 | 12374 | -/- | P16_23 | 25239 | 0 | 0 | 17469 | -/- | P32_12 | 26536 | 0 | 0 | 18153 | -/- |
| 8 COBAS negative samples | | | P8_46 | 26285 | 0 | 0 | 22379 | -/- | | | | | | | | | | | | |
| 8 COBAS negative samples | | | P8_47 | 25420 | 0 | 0 | 24184 | -/- | P16_24 | 26470 | 0 | 0 | 19399 | -/- | | | | | | |
| 8 COBAS negative samples | | | P8_48 | 26010 | 0 | 0 | 14457 | -/- | | | | | | | | | | | | |
| 8 COBAS negative samples | | | P8_49 | 24858 | 0 | 0 | 12808 | -/- | P16_25 | 25065 | 4 | 3 | 13301 | +/+ | P32_13 | 26706 | 2 | 6 | 12266 | +/+ |
| 79484 | 35.73 | 34.45 | P8_50 | 25900 | 7 | 7 | 13425 | +/+ | | | | | | | | | | | | |
| + 7 COBAS negative samples | | | | | | | | | | | | | | | | | | | | |
| 8 COBAS negative samples | | | P8_51 | 24696 | 0 | 0 | 16954 | -/- | P16_26 | 25356 | 0 | 0 | 11840 | -/- | | | | | | |
| 8 COBAS negative samples | | | P8_52 | 25416 | 0 | 0 | 4837 | -/- | | | | | | | | | | | | |
| 8 COBAS negative samples | | | P8_53 | 25625 | 0 | 0 | 12554 | -/- | P16_27 | 22665 | 0 | 0 | 9484 | -/- | P32_14 | 26044 | 2 | 1 | 6664 | +/+ |
| 8 COBAS negative samples | | | P8_54 | 23544 | 0 | 0 | 6078 | -/- | | | | | | | | | | | | |
| 83938 | 36.3 | 34.2 | P8_55 | 22501 | 5 | 3 | 2607 | +/+ | P16_28 | 20171 | 2 | 0 | 5581 | +/- | | | | | | |
| + 7 COBAS negative samples | | | | | | | | | | | | | | | | | | | | |
| 8 COBAS negative samples | | | P8_56 | 23211 | 0 | 0 | 7245 | -/- | | | | | | | | | | | | |

*Table S3: Results for A) individual reference RT-PCR testing; B) grouped testing by RT-dPCR for P8 extracts; C) grouped testing by RT-dPCR for P16 groups; and D) grouped testing by RT-dPCR for P32 groups. Highlighted in light green: results are in concordance (-/- or +/+). Highlighted in light orange: COBAS+/dPCR- discordance. Highlighted in light red: COBAS-/dPCR+ discordance. ND= "Not detected".*

## S4 – Detailed results of confirmatory testing for COBAS-/dPCR+ discordances

| Sample ID | COBAS reference | P8 extract number | dPCR+/ dPCR- | N cp / rnx | ORF1ab cp / rnx | IC cp / rnx | Altona E gene | Altona S gene | COBAS N | COBAS OFR |
|---|---|---|---|---|---|---|---|---|---|---|
| 51996 | *COBAS-* | 20 | *dPCR-* | 1 | 0 | 7 502 | *NT* | *NT* | *NT* | *NT* |
| 52019 | *COBAS-* | 20 | *dPCR-* | 0 | 0 | 25 944 | *NT* | *NT* | *NT* | *NT* |
| 52031 | *COBAS-* | 20 | *dPCR-* | 0 | 0 | 68 983 | *NT* | *NT* | *NT* | *NT* |
| 52035 | *COBAS-* | 20 | *dPCR-* | 0 | 0 | 28 215 | *NT* | *NT* | *NT* | *NT* |
| 52042 | *COBAS-* | 20 | *dPCR+* | 16 | 19 | a89 747 | 33 | 31.9 | ND | ND |
| 52047 | *COBAS-* | 20 | *dPCR-* | 0 | 0 | 122 825 | *NT* | *NT* | *NT* | *NT* |
| 52060 | *COBAS-* | 20 | *dPCR-* | 0 | 0 | 17 237 | *NT* | *NT* | *NT* | *NT* |
| 52062 | *COBAS-* | 20 | *dPCR-* | 0 | 0 | 105 711 | *NT* | *NT* | *NT* | *NT* |
| 56075 | *COBAS-* | 28 | *dPCR+* | 128 | 106 | 69 268 | 29 | 28.4 | 36.7 | ND |
| 56077 | *COBAS-* | 28 | *dPCR-* | 0 | 0 | 34 656 | *NT* | *NT* | *NT* | *NT* |
| 56083 | *COBAS-* | 28 | *dPCR-* | 0 | 0 | 71 410 | *NT* | *NT* | *NT* | *NT* |
| 56191 | *COBAS-* | 28 | *dPCR-* | 0 | 0 | 22 471 | *NT* | *NT* | *NT* | *NT* |
| 56211 | *COBAS-* | 28 | *dPCR-* | 0 | 0 | 21 185 | *NT* | *NT* | *NT* | *NT* |
| 56275 | *COBAS-* | 28 | *dPCR-* | 0 | 0 | 22 771 | *NT* | *NT* | *NT* | *NT* |
| 56303 | *COBAS-* | 28 | *dPCR-* | 0 | 0 | 56 797 | *NT* | *NT* | *NT* | *NT* |
| 56307 | *COBAS-* | 28 | *dPCR-* | 0 | 0 | 59 159 | *NT* | *NT* | *NT* | *NT* |
| 60241 | *COBAS-* | 39 | *dPCR-* | 0 | 0 | 46 988 | *NT* | *NT* | *NT* | *NT* |
| 60281 | *COBAS-* | 39 | *dPCR-* | 1 | 0 | 73 581 | *NT* | *NT* | *NT* | *NT* |
| 60310 | *COBAS-* | 39 | *dPCR-* | 0 | 0 | 11 420 | *NT* | *NT* | *NT* | *NT* |
| 60334 | *COBAS-* | 39 | *dPCR-* | 0 | 0 | 32 663 | *NT* | *NT* | *NT* | *NT* |
| 60345 | *COBAS-* | 39 | *dPCR-* | 0 | 0 | 81 004 | *NT* | *NT* | *NT* | *NT* |
| 60362 | *COBAS-* | 39 | *dPCR-* | 0 | 0 | 36 943 | *NT* | *NT* | *NT* | *NT* |
| 60389 | *COBAS-* | 39 | *dPCR-* | 0 | 0 | 825 | *NT* | *NT* | *NT* | *NT* |
| 60401 | *COBAS-* | 39 | *dPCR+* | 2 | 1 | 43 651 | ND | ND | ND | ND |

*Table S4: Results of the individual reassessment by both RT-dPCR and RT-PCR (Altona and Cobas®) for the samples from the 3 "COBAS-/dPCR+" discordant groups. NT = "Not tested". ND= "Not detected".*

### S5 - Absence of discordance in groups of 32

In this study, we did not observe any discordance for test results in groups of 32 when compared to the reference individual RT-PCR results. In particular and surprisingly, there were no qPCR+/dPCR- discordances although we would suspect an increased occurrence of such discordances as group size increases from the dilution effect of group testing.

Interestingly, the 2 samples that were associated with the 2 qPCR+/dPCR- discordances for groups of 16 (with Ct values of [36.7; *ND*] and [36.3; 34.2] respectively) are both P32 groups that contained one single positive qPCR+ sample. Consequently, these two samples were successfully detected by RT-dPCR in groups of 32 (P32_07 & P32_14) but not detected in the groups of 16 (P16_13 & P16_28).

In case of P32_07, the group of 32 is a combination of 2 P16 groups with opposite discordances: P16_13 (qPCR+/dPCR-) and P16_14 (qPCR-/dPCR+). It is likely that the RNA templates from the P16_14 group are those that were detected in the P32_07 group.

In the case of P32_14, the number of positive droplets observed in RT-dPCR is right above the threshold of 3 positive droplets (N=2; ORF1ab=1) while for the corresponding group of 16 (P16_28), the number of positive droplets is just below with 2 droplets (N=2; ORF1ab=0). In this case, statistical variations due to sampling error could explain the observation.

## S6 – Correlation between RT-dPCR measurements and Ct values

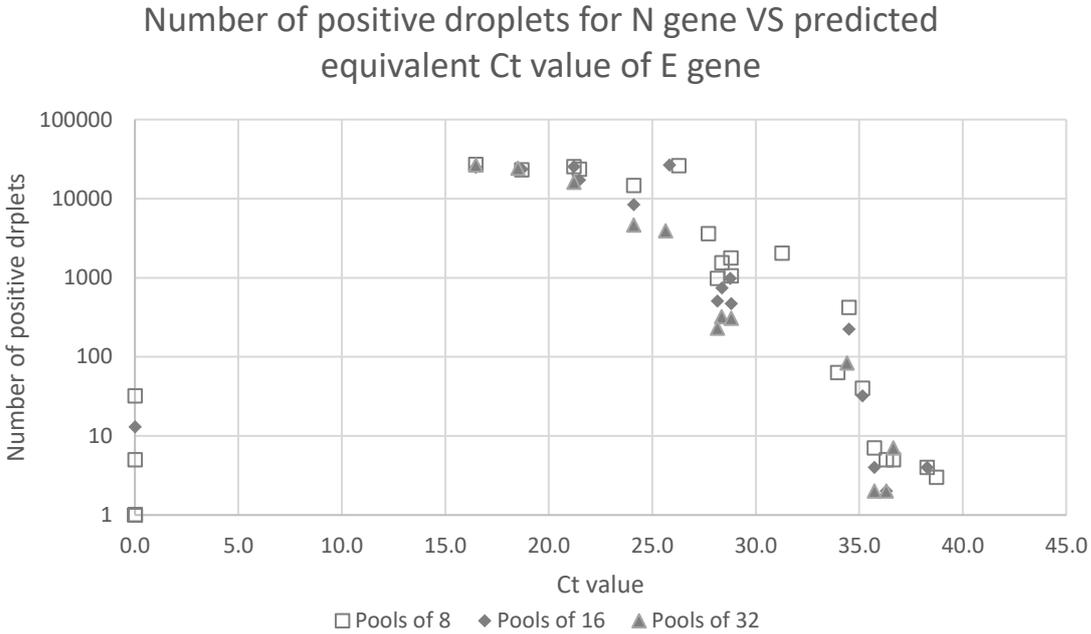

*Figure S6.1: Plot of the number of positive droplets measured by RT-dPCR for the N gene target in the groups of 8, 16 and 32 versus the predicted equivalent Ct value of the E gene of an RT-PCR measurement of the group using the Cobas® SARS-CoV-2 assay. The predicted equivalent Ct value of a group is defined as an average of the Ct values of the positive samples included in the group, taking into account the logarithmic scale of the Ct value.*

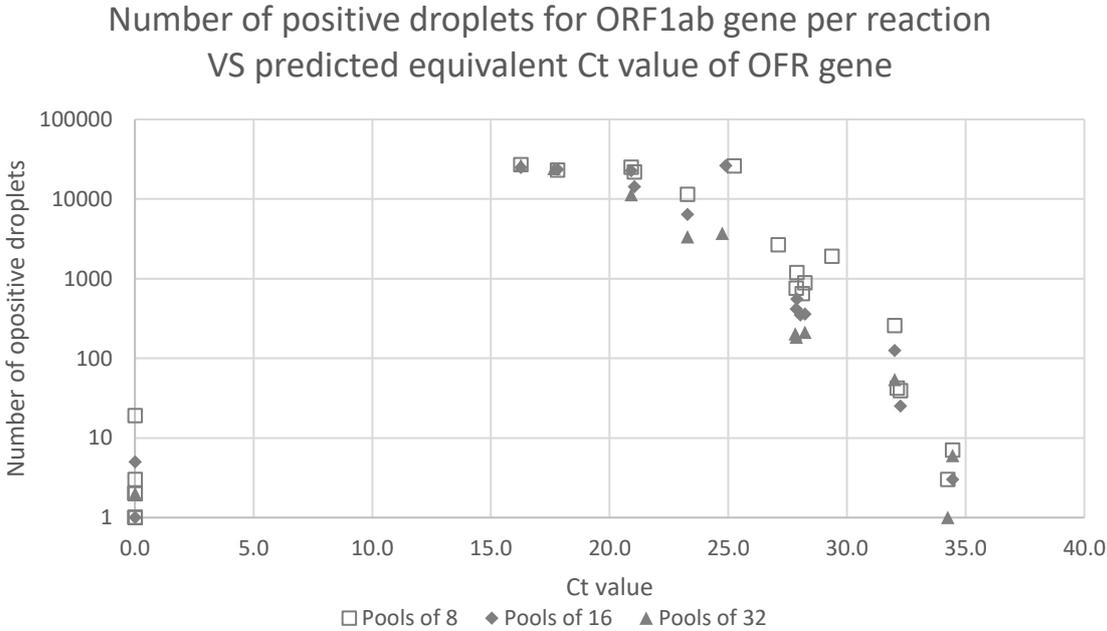

*Figure S6.2: Plot of the number of positive droplets measured by RT-dPCR for the ORF1ab gene target in the groups of 8, 16 and 32 versus the predicted equivalent Ct value of the ORF gene of an RT-PCR measurement of the group using the Cobas® SARS-CoV-2 assay. The predicted equivalent Ct value of a group is defined as an average of the Ct values of the positive samples included in the group, taking into account the logarithmic scale of the Ct value.*